\documentclass{optica-article}

\journal{opticajournal} 
\articletype{Research Article}

\usepackage{lineno}

\usepackage{amsmath}
\usepackage{placeins} 
\usepackage{float}    
\usepackage{flafter}  
\begin{document}

\title{Precision Hamiltonian Encoding in Full-Aperture Spatial Photonic Ising Machines}

\author{P. S. Karavelas,\authormark{1,2,3,*} D. Karanikolopoulos,\authormark{1} L. Mouchliadis,\authormark{2,4} A. K. Spiliotis,\authormark{1} N. L. Pitanios,\authormark{1,2,4} S. Gentilini,\authormark{5,6} D. Veraldi,\authormark{5} P. Charlesworth,\authormark{5} D. Pierangeli,\authormark{5} J. Sakellariou,\authormark{7} N. G. Berloff,\authormark{8} S. I. Tsintzos,\authormark{7} C. Conti,\authormark{5} and P. G. Savvidis\authormark{1,2,4}}

\address{\authormark{1}Department of Physics, School of Science, Westlake University, Hangzhou 310014, China\\
\authormark{2}Center for Quantum Science \& Technologies, FORTH-QuTech, Heraklion 70013, Crete, Greece\\
\authormark{3}Department of Physics, University of Crete, Heraklion 70013, Crete, Greece\\
\authormark{4}Department of Materials Science and Engineering, University of Crete, Heraklion 70013, Crete, Greece\\
\authormark{5}Department of Physics, Sapienza University, Piazzale Aldo Moro 5, 00185 Rome, Italy\\
\authormark{6}Institute for Complex Systems, National Research Council (ISC-CNR), Via dei Taurini 19, 00185 Rome, Italy\\
\authormark{7}QUBITECH, Thessalias 8, GR 15231 Chalandri, Athens, Greece\\
\authormark{8}Department of Applied Mathematics and Theoretical Physics, University of Cambridge, Cambridge CB3 0WA, United Kingdom}

\email{\authormark{*}pkaravelas@physics.uoc.gr}

\begin{abstract*}
Spatial photonic Ising machines (SPIMs) offer compact, room-temperature hardware with inherently parallel, energy-efficient, single-shot optical evaluation of the Ising Hamiltonian. However, accurate operation has been fundamentally limited by optical aberrations and non-uniform illumination, which corrupt phase-based spin encoding and distort coupling representation, forcing operation to a restricted spatial light modulator (SLM) region. Here we introduce a high-precision full-aperture calibration scheme that overcomes these constraints. By implementing wavefront retrieval and correction with $<\lambda/40$ accuracy, we restore faithful phase encoding across the entire SLM area. Furthermore, we introduce an interaction-normalization method, which compensates for amplitude curvature and enables uniform coupling representation. Together, these advances establish a full-aperture SPIM architecture whose faithful Hamiltonian encoding is a prerequisite for photonic Ising computation.
\end{abstract*}

\section{Introduction}
The Ising model is a cornerstone of statistical mechanics and a powerful framework for solving combinatorial optimization problems by mapping them to the search for low-energy configurations of an Ising Hamiltonian \cite{ref1,ref2,ref3}. Many NP-hard problems can be formulated as Quadratic Unconstrained Binary Optimization (QUBO) tasks, turning combinatorial search into an energy-minimization problem well suited for physics-inspired hardware acceleration \cite{ref2}. Ising machines exploit this principle by allowing a physical system to evolve toward low-energy states, providing candidate solutions to optimization problems \cite{ref3}. This capability has led to wide interest across applications including scheduling \cite{ref4}, materials science \cite{ref5}, protein folding \cite{ref6}, chemical design \cite{ref7}, finance \cite{ref8}, traffic \cite{ref9}, and communication \cite{ref10}.

Among existing approaches, photonic Ising machines \cite{ref5,ref11,ref12,ref13,ref14,ref15,ref16,ref17,ref18} are particularly promising because they implement spins and interactions using optical fields, enabling massive parallelism and low latency. Optical propagation can perform global summations or inner products in a single step, offering the potential for high throughput and improved energy efficiency compared with electronic systems \cite{ref19,ref20,ref21,ref22}. In this context, spatial photonic Ising machines encode spins in the spatial degrees of freedom of a coherent optical field and use propagation and detection to evaluate Hamiltonian terms in parallel. Typically, spins are implemented by assigning each lattice site to a spatial region of a spatial light modulator that applies discrete phase values (commonly $0$ or $\pi$). After the SLM, a lens performs a Fourier transform so that the far-field intensity measured by a camera becomes a quadratic function of the output field, computing weighted sums of pairwise spin products through interference. This enables single-shot optical evaluation of the Ising energy with high resolution and large dynamic range, making SPIMs attractive for NP-hard tasks such as max-cut, spin-glass ground-state search, graph partitioning, and other QUBO formulations. A central challenge, however, is the encoding of arbitrary dense coupling matrices $J$, which are not addressed by the initial SPIM approach \cite{ref12}, restricted only to Mattis-type Ising Hamiltonians \cite{ref23}. Implementing general interactions therefore requires decomposition strategies---such as eigen-decomposition or low-rank approximations---followed by space-, wavelength-, or time-division multiplexing \cite{ref24,ref25,ref26,ref27,ref28,ref29}. Another approach employs an additional ancilla spin that selects the integrated information on the focal plane, allowing arbitrary interactions to be encoded \cite{ref30}.

Importantly, SPIM performance is strongly limited by wavefront distortion and amplitude non-uniformity, both of which compromise accurate Hamiltonian encoding. Aberrations arise inevitably from imperfect optical elements and distort the propagated wavefront, leading to incorrect phase mappings and errors in effective spin interactions. Similarly, a uniform amplitude profile is required for precise interaction encoding \cite{ref12}. For this reason, many SPIM implementations restrict operation to a central SLM region where aberrations are minimized \cite{ref12,ref25,ref31,ref32,ref33,ref34,ref35,ref36}. However, enlarging the usable area amplifies distortion effects and restricts scalability. Even small wavefront deviations can cause incorrect mapping, preventing reliable use of the full SLM aperture.

Aberration correction for SPIMs has been explored previously \cite{ref37}. Here we introduce a full-aperture calibration that corrects both wavefront distortion and amplitude non-uniformity across the entire SLM, using an interferometer-free, common-path scheme that requires no external reference beam or active stabilization and is therefore insensitive to thermal and vibrational drift. We implement a high-precision wavefront retrieval that compensates optical aberrations to a root-mean-square error below $\lambda/40$ \cite{ref38,ref39}, restoring faithful phase encoding over the full aperture. We then address a second, equally critical effect: beam expansion imprints a non-uniform Gaussian intensity profile that acts as a spatially dependent weight and biases the encoded couplings. We remove it by normalizing the encoded interactions to the retrieved amplitude profile, ensuring uniform coupling and enabling full utilization of the SLM surface.

\section{Results}
\subsection{Full-aperture phase retrieval and correction}

\begin{figure*}[t!]
\centering\includegraphics[width=\linewidth]{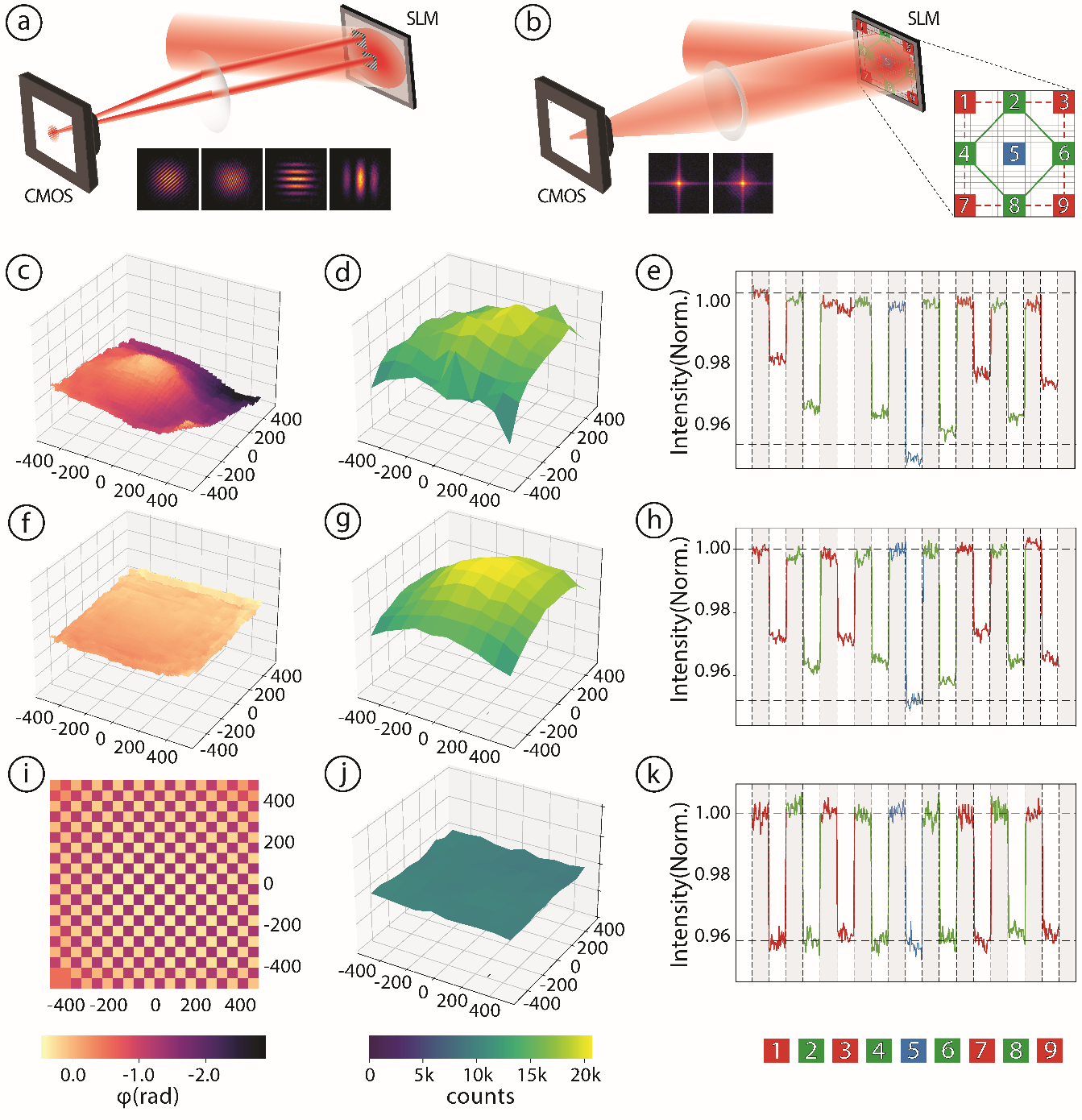}
\caption{Wavefront retrieval and correction methods. Schematics of (a) the phase retrieval method, where the inset shows four representative interferograms, and (b) the field-intensity retrieval process, where the inset shows the diffraction patterns corresponding to the fully polarized and the not-fully-polarized spin configurations. Phase correction: (c) retrieved phase variation before and (f) after phase correction. Intensity correction: (d) reconstructed optical field intensity distribution across the SLM before any correction, (g) after phase correction, and (j) after both phase correction and amplitude flattening. (i) The checkerboard pattern on the SLM implementing the amplitude flattening. Spin-flip diagnostic test: (e) before corrections, (h) after only phase correction, and (k) with both phase correction and amplitude flattening.}
\end{figure*}
The comprehensive wavefront correction scheme is detailed in Fig.~1, illustrating the retrieval procedure adapted from Schroff \emph{et al.} \cite{ref39}. This common-path approach extracts the local phase difference across the SLM plane by interfering a fixed reference region with a scanning sampled region, producing the self-referenced interferograms shown in Fig.~1a. This method allows for high-precision mapping of phase deviations without the need for an external reference beam. The phase values $\delta\varphi$, extracted by fitting the interferograms between a sampled region and the reference, are used to construct a corrective phase mask that compensates for optical distortions (Fig.~1c), effectively restoring a flat laser phase front (Fig.~1f). This correction achieves high precision, with a root-mean-square accuracy of $\lambda/40$ and a peak-to-valley deviation of approximately $\lambda/6$. Implementing this correction ensures that each phase-encoded spin state represents the intended Ising model configuration.

The performance of the SPIM critically depends on its ability to detect single spin-flip events and accurately quantify their contribution to the total Ising energy, which is measured as the optical intensity on the focal plane \cite{ref24,ref25}. The influence of a single spin flip on the recorded Ising energy can be evaluated by first measuring the intensity when all spins are uniformly polarized in one direction, and then measuring the intensity after flipping a selected spin. The corresponding diffraction patterns on the camera are shown in the inset of Fig.~1b. By subtracting the intensity in the single-flip configuration from the fully polarized intensity, the individual contribution of the flipped spin to the Ising energy of the Mattis-model ferromagnetic interaction ($J_{ij}=\xi_i\xi_j$) is expected to approximately match the optical field intensity at the position of the flipped spin. By systematically scanning the entire spin lattice, this method allows for the experimental retrieval of each spin's contribution to the Ising energy and effectively reconstructs the optical field intensity distribution across the SLM (Figs.~1d, g, j). Because each single-flip signal measures the local field at that macropixel, this reconstruction is not merely diagnostic: the retrieved profile serves directly as the interaction-normalization map, so the SPIM builds its own correction from its own readout (Appendix~B).

Prior to the use of the correction mask (Fig.~1c), the reconstructed intensity does not follow the anticipated Gaussian distribution but rather a distorted profile (Fig.~1d). The impact of this distorted wavefront is further illustrated in Fig.~1e through a diagnostic spin-flip experiment. In this test, a subset of nine representative spins, schematically indicated in Fig.~1b as differently colored SLM pixels, are sequentially flipped. For each spin-flip event, the corresponding contribution to the overall intensity is measured. The results reveal that the distorted wavefront introduces a local phase offset at the spin-encoding sites, leading to spins that are either nonresponsive (Fig.~1e) or do not contribute in accordance with the encoded interaction coupling. Importantly, this effect scales with the physical SLM size, presenting a fundamental limitation to the scalability of SPIMs.

We proceed to calibrate the SPIM implementation by first applying the calculated correction mask on the SLM to compensate for the optical aberrations. This results in the desired flat phase profile illustrated in Fig.~1f and in an intensity distribution shown in Fig.~1g where the Gaussian profile is restored. In addition, a fundamental prerequisite for reliably mapping spin-system problems onto SPIMs is the accurate implementation of inter-spin interactions. To enable full-area utilization of the SLM, we analyze the measured intensity profile in Fig.~1g and observe pronounced non-uniformity across the illuminated region. Conceptually, the inherent Gaussian beam profile (Fig.~1g) can be understood as introducing an additional, unintended layer of interactions embedded within the beam itself. Building on the work of Fang \emph{et al.} \cite{ref32,ref33,ref40}, who demonstrated that both spin configurations and inter-spin interactions can be encoded using a single SLM, improving upon earlier approaches requiring two separate SLMs \cite{ref12,ref31,ref34,ref41,ref42}, we compensate the non-uniform Gaussian intensity profile using a phase-only SLM. Guided by this encoding principle, we design a suitable transformation that ensures the resulting amplitude after interference matches the minimum amplitude of the initial field (see Appendix~B for the full derivation). This transformation can be generalized as an interaction-normalization method that removes the additional interaction layer caused by the non-uniformity of the optical beam. As demonstrated in Fig.~1j, the encoding scheme produces the expected uniform intensity profile for ferromagnetic (FM) interactions ($J_{ij}>0$), where all spins contribute equally. The improvement is further verified by spin-flip diagnostics: while the previous implementation showed variations in spin-flip contributions as large as 45\% (Fig.~1h), the optimized system reduces this discrepancy to less than 10\% (Fig.~1k). The calibration of the SPIM is completed with this final step, ensuring accurate and precise encoding of Ising spins and their interactions. This initialization is crucial for implementing photonic simulations across the full SLM area.

\subsection{Mattis spin-glass validation}

\begin{figure*}[t!]
\centering\includegraphics[width=\linewidth]{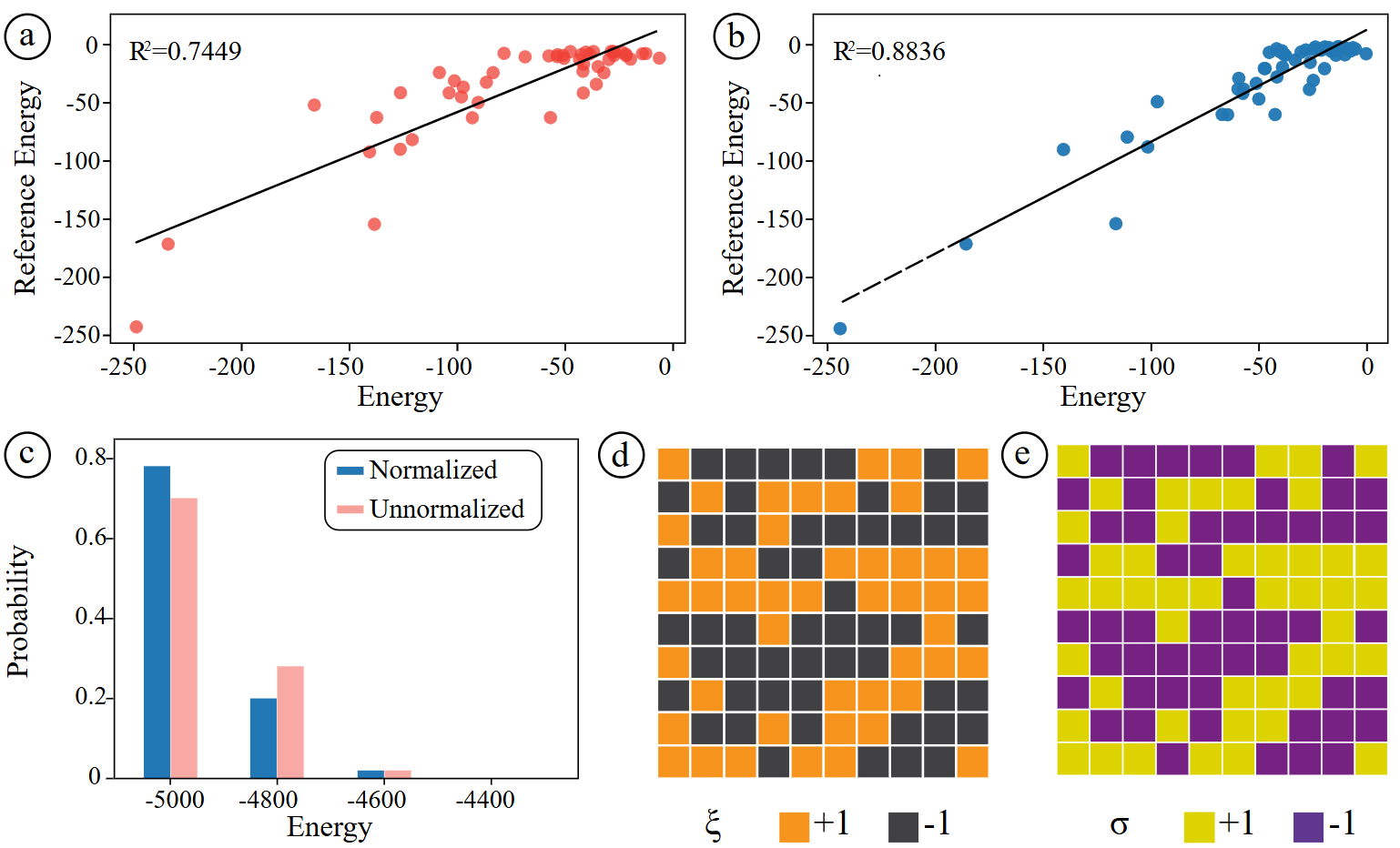}
\caption{Comparison of normalized vs.\ un-normalized interactions. (a,b) Correlation between experimental intensities and the analytical Mattis Hamiltonian before (a) and after (b) interaction normalization. (c) Comparison of ground-state search experiments between the normalized and un-normalized cases over 50 experiments of 1000 Metropolis--Hastings steps at $T=0$. (d) Interaction coupling for the Mattis spin-glass problem. (e) Calculated ground state of the system, consistent with the vector $\xi_i$.}
\end{figure*}
In a recent study, Sakellariou \emph{et al.} \cite{ref30} proposed a calibration approach that correlates experimentally measured with theoretically predicted energies. This approach involves computing the energy of different spin configurations using both the SPIM and the analytical Sherrington--Kirkpatrick (SK) model Hamiltonian \cite{ref43}. A linear regression is then used to correlate the results from these two methods. For the SPIM to accurately simulate the encoded problem, a linear relation between the two energy sets is expected. This is consistent with existing literature, provided that the wavefront (both phase and amplitude) is relatively flat, to ensure accurate problem encoding.

Our results show that achieving such uniformity across the entire SLM region is an essential initial step. Phase correction enables the calibration of the system's capacity for spin-configuration representation, while interaction normalization ensures precise encoding of spin couplings. To demonstrate the effect of these corrections on the relationship between the two energies, we conducted both numerical simulations and experimental measurements. To evaluate the influence of the normalized encoding, we solved a Mattis spin-glass model with interaction couplings $J_{ij}=\xi_i\xi_j$.

The selection of this model is based on its simplicity in evaluating ground-state energies. In Mattis-type interaction systems, the expected ground-state spin configuration coincides with the vector $\xi_i$ \cite{ref1,ref12} or its inverse, and can therefore be directly verified (Fig.~2d, e). The objective of this study is to assess the system's ability to accurately encode interactions, based on both energy calibration and convergence towards the ground state. As illustrated in Fig.~2a, the correlation between the analytical Mattis Hamiltonian and the experimentally retrieved energies using unnormalized interactions demonstrates a substantial mismatch. As demonstrated in Fig.~2b, the application of interaction normalization results in enhanced agreement, thereby improving the system's Hamiltonian representation. Subsequent to the calibration procedure, the experiment was repeated 50 times using the Metropolis--Hastings algorithm with 1000 steps at $T=0$. The normalization process significantly improves the performance of the ground-state search (see Fig.~2c). For the normalized case, we observe an improved agreement with $R^2=0.88$. When interactions are improperly normalized and in the absence of compensation, the Ising machine is significantly affected by experimental artifacts.

\subsection{Sampled spin lattices}

\begin{figure*}[t!]
\centering\includegraphics[width=\linewidth]{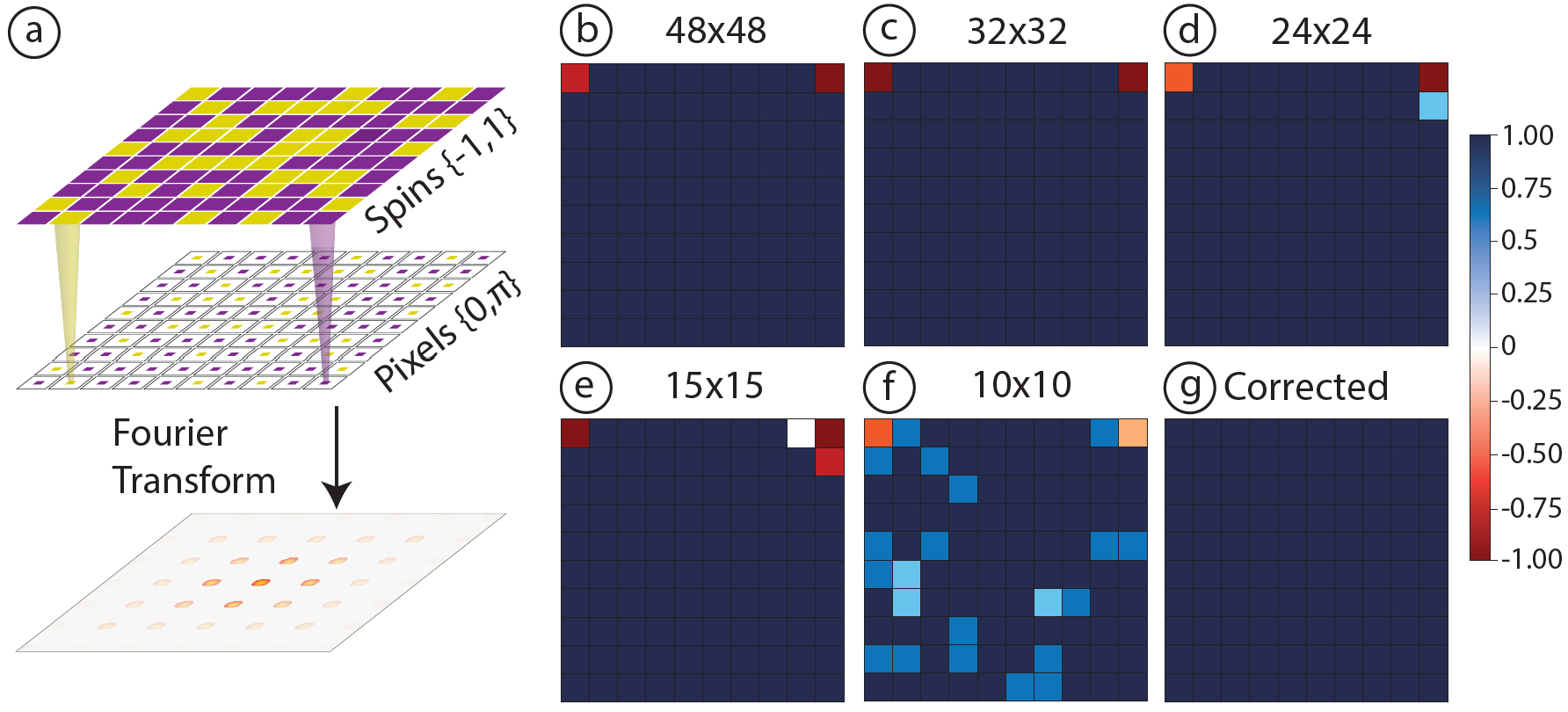}
\caption{Experimental demonstration of sampled spin lattices, including mean spin configurations. (a) Schematic of the experimental procedure for generating sampled spin lattices. (b)--(f) Mean spin configurations from ten FM ground-state search runs without phase correction, for five different MP sizes, after the three post-processing steps. (g) Mean spin configurations from ten FM ground-state search runs with phase correction for all MP sizes.}
\end{figure*}
We proceed to show that precise wavefront correction is a key enabler of SPIM scalability, towards accurate encoding over the full SLM aperture. To achieve this, we conduct two distinct experimental procedures. The first approach tests FM interactions by gradually increasing the system size and observing the effects on configuration, energy, magnetization, and stability. However, as the system size increases, this method proves extremely time-consuming, prompting the design of an alternative experiment with the same constraints but greater efficiency. This modified experiment utilizes sampled macropixels distributed across the entire SLM region, as shown in Fig.~3a. By covering the full SLM surface, phase-distortion and amplitude-curvature effects can be tracked by focusing only on a subset of representative spins. The scalability test is conducted by keeping a constant working area on the SLM ($960\times960$ pixels) while decreasing the macropixel size. This method enables the experimental realization of progressively larger spin lattices, aiming to determine the minimum macropixel size required to solve the ferromagnetic interaction problem. We test macropixels with side lengths of 48, 32, 24, 15, and 10 pixels, which correspond to mimicking lattices of 400, 900, 1600, 4096, and 9216 spins, respectively. Note that in these spin-lattice simulations, we employ a subset of 100 spins equally distributed across the SLM region while we do not use the rest of the SLM area. This sparse sampling approach ensures coverage of the entire active SLM area while maintaining computational efficiency, hence we use the term sampled spin lattices.

Following each experimental convergence event, we perform three post-processing steps: (i) domain standardization: we manually normalize the resulting spin configuration by setting the largest domain equal to $+1$ and all remaining spins (or smaller domains) equal to $-1$; (ii) ensemble averaging: we repeat this process across 10 independent convergence trials; (iii) statistical analysis: the final mean spin configuration is computed as the ensemble average of all standardized configurations. This post-processing ensures consistent comparison across trials while preserving the dominant domain structure in our statistical analysis.

In Figs.~3b--3f, we show the average spin polarization in the final-state configuration for five different macropixel (MP) sizes, before applying the SPIM calibration (phase correction and amplitude normalization). Without SPIM calibration, the final state deviates from the fully polarized FM ground state for all MP sizes. This deviation becomes more pronounced as the MP size decreases and correspondingly the spin-lattice size increases, suggesting that while the Metropolis algorithm drives the system toward its ground state, some spins---particularly at the edges of the SLM, where optical aberrations are strongest---fail to align with the dominant orientation. In contrast, after correcting for optical aberrations, the system converges to the theoretically expected FM ground state, with all spins uniformly polarized, regardless of MP size (Fig.~3g). This confirms that SPIM calibration ensures a faithful representation of the Ising Hamiltonian. Notably, for the smallest MP size tested ($10\times10$ pixels), this approach combined with averaging could emulate spin lattices of up to ${\sim}10{,}000$ spins, demonstrating a clear prerequisite for scalable SPIMs.

\subsection{Magnetization convergence}

To further support the effect of the SPIM calibration, we focus on the magnetization as the key observable, tracking its convergence towards the FM ground state, which corresponds to a fully polarized spin system with a magnetization value of unity. In Fig.~4a we present the magnetization for three different lattice sizes of 100, 400, and 900 spins using the full-lattice approach. Upon convergence to the FM ground state, the system exhibits magnetization deviating from unity, with this deviation becoming increasingly larger as the system size increases. This indicates that although the Metropolis algorithm converges, the final-state spin configuration does not correspond to that of a FM ground state with all spins polarized. In contrast, when the phase-correction mask is applied, the magnetization converges to unity for all three lattices as expected from the theoretical FM ground-state solution of the Ising model (Fig.~4b).

\begin{figure*}[t!]
\centering\includegraphics[width=\linewidth]{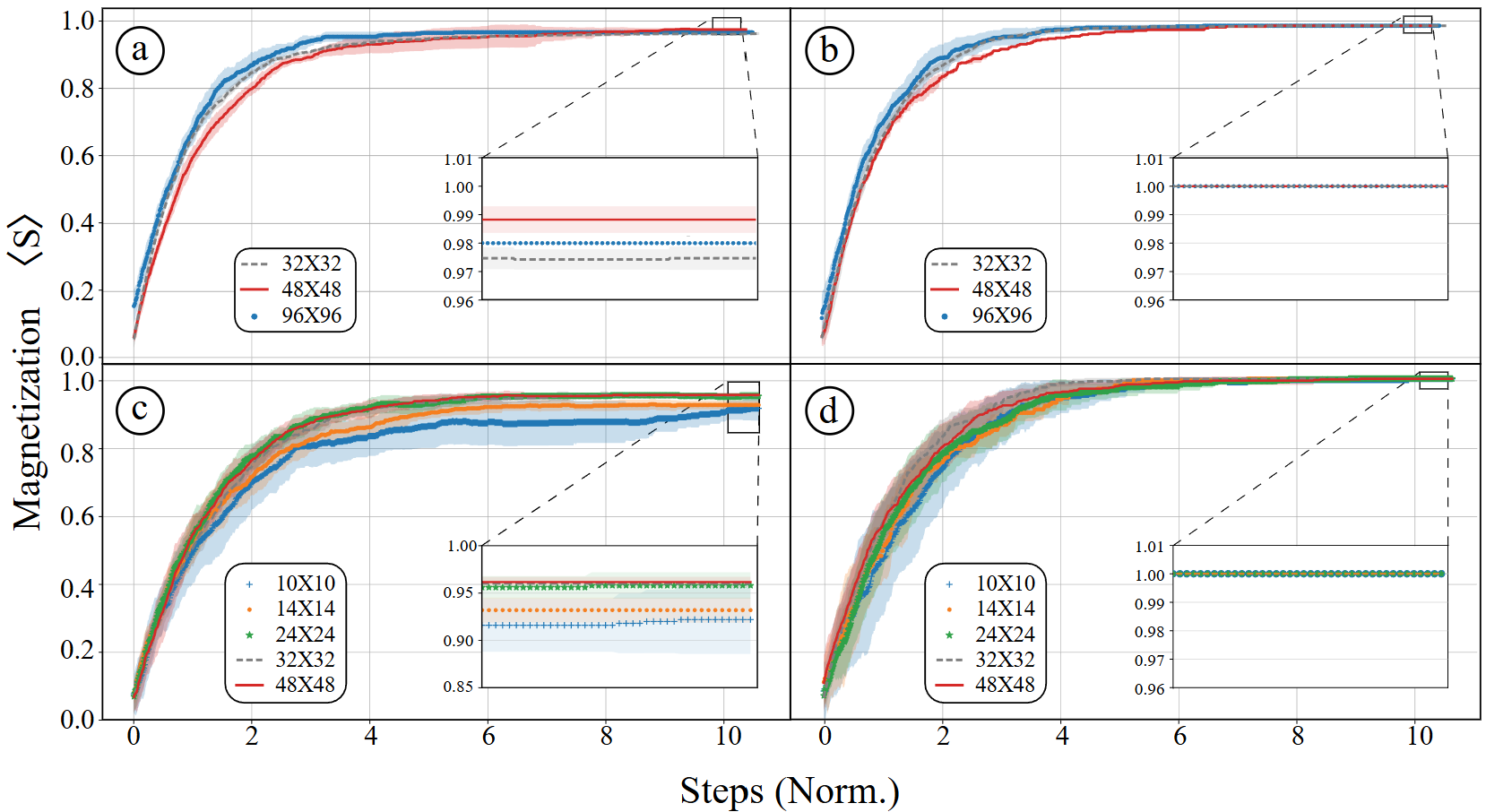}
\caption{Effect of phase correction on magnetization with increasing lattice size, using two experimental procedures. (a) Magnetization for three lattice sizes---100, 400, and 900 spins---before phase correction. (b) Magnetization for the same lattice sizes after applying phase correction. In both cases, convergence steps are normalized by the number of spins. In (c) and (d) we similarly track the magnetization before and after phase correction, respectively, this time by reducing the macropixel lattice-mask size while keeping the same ROI and spin-lattice size, mimicking spin lattices with sizes ranging from 400 to 9216 spins.}
\end{figure*}

Similarly, Figs.~4c and 4d display the magnetization before and after phase correction, respectively, for the sampled spin-lattice method. In the uncorrected case, the final-state magnetization fails to converge to unity as the macropixel size decreases. However, with phase correction, the system achieves complete spin polarization even for the largest spin lattices examined.

\FloatBarrier
\section{Conclusion}
The performance of spatial photonic Ising machines critically depends on accurate mapping between the Ising-model spin values and interactions and the phases encoded on the SLM. Aberrations, which arise from imperfections in optical components or misalignments, introduce deviations in the phase profile, leading to errors in the representation of spin states and coupling interactions. These errors manifest as a mismatch between the physical system and the mathematical model, reducing the accuracy of the solution. Compensating for aberrations is therefore crucial as it directly impacts the fidelity of the Ising-model representation on the SLM. As demonstrated by our analysis, a phase retrieval and consequent phase correction along with amplitude normalization to ensure constant phase and intensity across the entire SLM area is not merely a means to increase performance in terms of speed and computational accuracy but rather a \emph{sine qua non} procedure for correctly implementing the Ising Hamiltonian on a SPIM. Especially when the issue of scalability is considered, the phase and amplitude calibration proposed in this study renders itself an indispensable initialization step of next-generation SPIMs.

Our methodology bridges the gap between theoretical Ising models and physical implementations of SPIMs. By demonstrating that optical imperfections can be systematically corrected, we resolve a key limitation that has hindered the reliable application of SPIMs to large-scale problems. This opens the way for further advancements in scalable SPIMs. Compact photonic Ising machines \cite{ref25,ref30} may adopt our techniques to enhance portability without sacrificing accuracy. While our approach achieves remarkable precision, several challenges remain to achieve fully scalable SPIMs. For instance, thermally induced wavefront drift or SLM aging may necessitate adaptive, real-time correction algorithms designed to address dynamic aberrations. Additionally, enhancing energy-measurement efficiency could improve the ability of the system to decide whether a spin flip should be accepted or not.

In conclusion, by rigorously addressing phase and amplitude non-idealities, we have established a precision, interferometer-free calibration protocol that ensures faithful full-aperture Hamiltonian encoding---a prerequisite for scalable SPIMs. Our results confirm that optical Ising machines can faithfully emulate spin systems when physical encodings are precisely controlled, a prerequisite for their adoption in real-world optimization tasks. Future work should explore these techniques in non-ideal environments and for non-uniform interaction topologies.

\section*{Appendix A: SPIM setup}
The experimental setup is shown in Fig.~5. Our base laser system is a CW laser (Toptica iBeam Smart) with a central wavelength of $\lambda=811$~nm. The beam is initially coupled to a single-mode fiber, which serves as a spatial filter that creates a clean Gaussian wavefront. At the fiber exit, the beam is collimated by an $f=36$~mm fiber collimator and set to horizontal polarization by a polarizing beam splitter. A combination of $f_1=-50$~mm and $f_2=400$~mm lenses is used to expand the beam before we direct it to the SLM (Hamamatsu X13138). This ensures that the full SLM screen is covered. With the use of a blazed grating phase on the SLM, the beam is then directed to lens L3 ($f_3=300$~mm), where it is focused on the CMOS camera (Hamamatsu ORCA qCMOS) for detection.

\begin{figure}[htbp]
\centering\includegraphics[width=\linewidth]{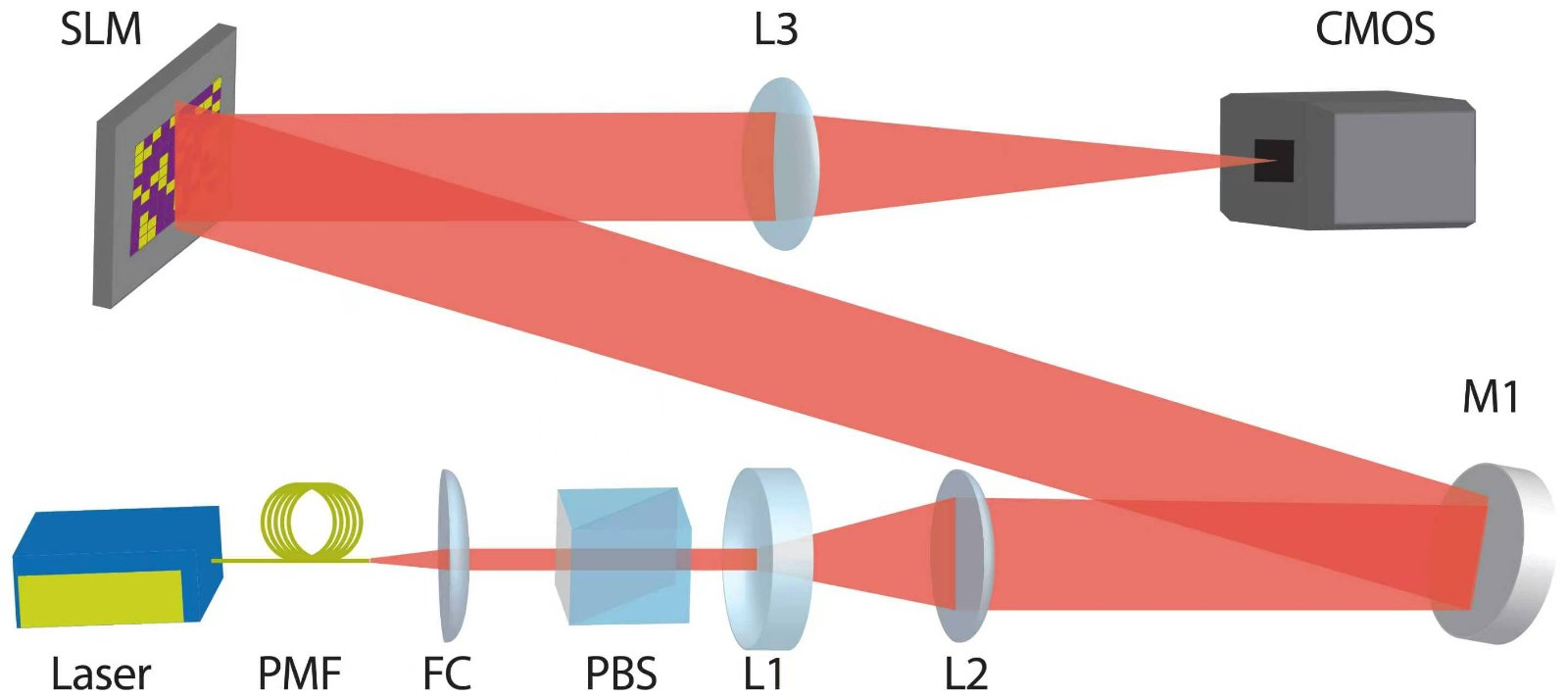}
\caption{Schematic of the experimental setup.}
\end{figure}

\section*{Appendix B: Interaction-normalization derivation}
A uniform amplitude profile across the SLM is a prerequisite for encoding a constant interaction. Because beam expansion imprints a Gaussian amplitude profile $A(x,y)$, the full-aperture field acts as an unintended, spatially dependent coupling that biases the encoded Ising interactions. We remove this bias by normalizing the encoded interactions to the retrieved amplitude profile, as follows. Beginning from a complex field with spatially varying amplitude and phase,
\begin{equation}
A(x,y)\,\exp\!\left(i\varphi(x,y)\right),\tag{B1}
\end{equation}
we synthesize two auxiliary fields of equal amplitude but distinct phases whose interference yields an intensity equal to the minimum intensity of the initial field,
\begin{equation}
C(x,y)\,\exp\!\left(i\lambda(x,y)\right)+C(x,y)\,\exp\!\left(i\mu(x,y)\right),\tag{B2}
\end{equation}
with
\begin{align}
\lambda(x,y)&=\varphi(x,y)+\cos^{-1}\!\big(A_{\min}/A(x,y)\big),\tag{B3a}\\
\mu(x,y)&=\varphi(x,y)-\cos^{-1}\!\big(A_{\min}/A(x,y)\big),\tag{B3b}
\end{align}
and
\begin{equation}
C(x,y)=A(x,y)/2.\tag{B4}
\end{equation}
The two phases are multiplexed onto a single phase-only element $\beta(x,y)$ using complementary binary (checkerboard) gratings $M_1,M_2$ satisfying $M_1+M_2=1$,
\begin{equation}
M_1\,\exp\!\left(i\lambda(x,y)\right)+M_2\,\exp\!\left(i\mu(x,y)\right)=\exp\!\left(i\beta(x,y)\right),\tag{B5}
\end{equation}
\begin{equation}
\beta(x,y)=M_1\,\lambda(x,y)+M_2\,\mu(x,y).\tag{B6}
\end{equation}
This redistributes the excess central intensity away from the region of interest so that every site contributes equally, matching the minimum-amplitude site and thereby encoding a constant ferromagnetic interaction. For non-constant couplings, the factor $A_{\min}/A(x,y)$ is applied as a correction coefficient multiplying the intended interaction, which eliminates the intrinsic interaction layer embedded within the beam. The construction follows the single-element phase-only encoding of Refs.~\cite{ref40} and \cite{ref32}.

The amplitude profile $A(x,y)$ entering Eqs.~(B1)--(B6) is obtained \emph{in situ} from the single-spin-flip diagnostic, so no separate amplitude measurement is required. Starting from the fully polarized ferromagnetic state, flipping spin $i$ changes the focal-plane ($k=0$) intensity by $\Delta I_i$, which for a ferromagnetic encoding is proportional to the local optical intensity at that macropixel, $\Delta I_i \propto A_i^2$. Scanning all spins therefore reconstructs the amplitude map $A_i \propto \sqrt{\Delta I_i}$ across the aperture (Figs.~1d, g, j). The interaction-normalization weight for site $i$ follows directly as the inverse amplitude ratio $A_{\min}/A_i = \sqrt{\Delta I_{\min}/\Delta I_i}$, identifying the measured profile with $A(x,y)$ in Eq.~(B3). The same diagnostic that reveals the non-uniform coupling thus supplies its correction.

\section*{Appendix C: Noise on the detection plane}
Single-spin-flip detection is ultimately noise-limited for small macropixels. Figure~6 shows the spin-flip signal as the macropixel size is progressively reduced, together with the effect of ensemble averaging. Averaging suppresses the detection noise and restores reliable single-spin-flip discrimination, supporting the averaging protocol used in the main text.

\begin{figure}[H]
\centering\includegraphics[width=\linewidth]{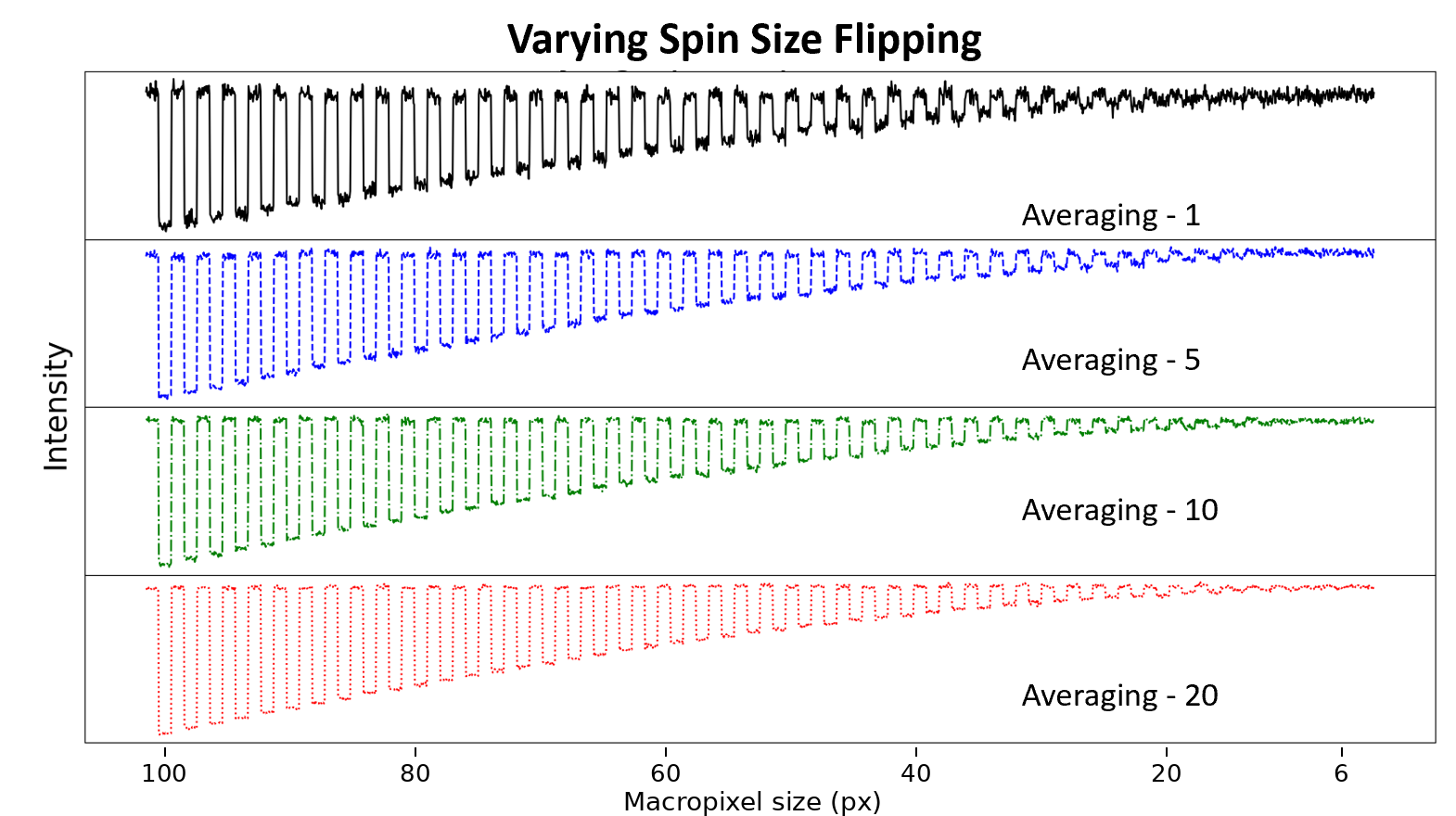}
\caption{Noise on the detection plane. Spin-flip signal as the macropixel size is reduced from $100\times100$ to $6\times6$ (step $-2$), for averaging over 1, 5, 10, and 20 frames; averaging suppresses noise and restores single-flip detection.}
\end{figure}

\FloatBarrier
\begin{backmatter}
\bmsection{Funding}
Program for Quantum Science and Technology (2023Zd0300300); Zhejiang Provincial Natural Science Foundation (Xhd24A2401); Horizon EIC 2022 Pathfinder Challenges 01 HEISINGBERG (101114978); European Union NextGenerationEU--Greece 2.0 National Recovery and Resilience Plan QuInPhoS ($\Upsilon\Pi$3TA-0561193).

\bmsection{Acknowledgment}
A.~K.~S.\ and P.~G.~S.\ acknowledge the support of the innovation Program for Quantum Science and Technology (2023Zd0300300) and the Zhejiang Provincial Natural Science Foundation (Xhd24A2401). Financial support from the Horizon EIC-2022 Pathfinder Challenges-01 HEISINGBERG project 101114978 is acknowledged. This work was supported by the European Union NextGenerationEU through the Greece 2.0 National Recovery and Resilience Plan, under the Call SUB.1.1 Clusters of Research Excellence (CREs), under Project QuInPhoS.

\bmsection{Disclosures}
The authors declare no conflicts of interest.

\bmsection{Data availability}
Data underlying the results presented in this paper are not publicly available at this time but may be obtained from the authors upon reasonable request.
\end{backmatter}


\end{document}